\documentclass[twocolumn,superscriptaddress,footinbib,aps,prl]{revtex4-1}
\usepackage[english]{babel}
\usepackage{epstopdf}
\usepackage{float}
\usepackage{enumerate}
\usepackage{hyperref}

\usepackage{amsmath} % AMS Math Package
\usepackage{cancel}
\usepackage{amsthm} % Theorem Formatting
\usepackage{amssymb}	% Math symbols such as \mathbb
\usepackage{graphicx} % Allows for eps images
\usepackage{empheq}
\usepackage[absolute,overlay]{textpos}
\usepackage{xcolor}

\begin{document}

\title{{\color{blue} Optimized Observable Readout from Single-shot Images of Ultracold Atoms\\ via Machine Learning}}
\date{\today}

\author{Axel U. J. Lode}
\email{auj.lode@gmail.com}
\affiliation{Institute of Physics, Albert-Ludwig University of Freiburg, Hermann-Herder-Strasse 3, 79104 Freiburg, Germany}
\author{Rui Lin}
%\email{molignini@itp.phys.ethz.ch}
\affiliation{Institute for Theoretical Physics, ETH Z\"{u}rich, 8093 Zurich, Switzerland}
\author{Miriam B\"{u}ttner}
\affiliation{Institute of Physics, Albert-Ludwig University of Freiburg, Hermann-Herder-Strasse 3, 79104 Freiburg, Germany}
\author{Luca Papariello}
\affiliation{Research Studio Data Science, RSA FG, Thurngasse 8/16, 1090 Wien, Austria}
\author{Camille L\'{e}v\^{e}que}
\affiliation{Vienna Center for Quantum Science and Technology,
Atominstitut, TU Wien, Stadionallee 2, 1020 Vienna, Austria}
\affiliation{Wolfgang Pauli Institute c/o Faculty of Mathematics,
University of Vienna, Oskar-Morgenstern Platz 1, 1090 Vienna, Austria}
\author{R. Chitra}
\affiliation{Institute for Theoretical Physics, ETH Z\"{u}rich, 8093 Zurich, Switzerland}
\author{Marios C. Tsatsos}
\affiliation{Honest AI Ltd., 65 Goswell Road, London EC1V 7EN, United Kingdom}
\author{Dieter Jaksch}
\affiliation{Clarendon Laboratory, University of Oxford, Oxford OX1 3PU, United Kingdom}
\author{Paolo Molignini}
\email{phys2088@ox.ac.uk}
\affiliation{Clarendon Laboratory, University of Oxford, Oxford OX1 3PU, United Kingdom}

%%%%%%%%%%%%%%%%%%%%%%%%%%%%%%%%%%%%%%%%%%%%%%
%%%%               		   	          ABSTRACT              					%%%%
%%%%%%%%%%%%%%%%%%%%%%%%%%%%%%%%%%%%%%%%%%%%%%
\begin{abstract}
Single-shot images are the standard readout of experiments with
ultracold atoms -- the tarnished looking glass into their many-body physics. 
The efficient extraction of observables
from single-shot images is thus crucial. Here, we demonstrate how
artificial neural networks can optimize this extraction.
In contrast to standard averaging approaches, machine learning allows both
one- and two-particle densities to be accurately obtained from a
drastically reduced number of single-shot images. 
Quantum fluctuations and correlations are directly harnessed to obtain physical
observables for bosons in a tilted double-well potential at an unprecedented accuracy. Strikingly, machine learning also
enables a reliable extraction of momentum-space observables from
real-space single-shot images and vice versa. This obviates the need for a reconfiguration of the experimental setup between in-situ and time-of-flight imaging, thus potentially granting an outstanding reduction in resources.
\end{abstract}
%%%%%%%%%%%%%%%%%%%%%%%%%%%%%%%%%%%%%%%%%%%%%%

\maketitle

%%%%%%%%%%%%%%%%%%%%%%%%%%%%%%%%%%%%%%%%%%%%%%
%%%%  INTRODUCTION                        %%%%
%%%%%%%%%%%%%%%%%%%%%%%%%%%%%%%%%%%%%%%%%%%%%%

Ultracold atoms are used as remarkably flexible analog quantum simulators of otherwise hardly accessible quantum many-body states~\cite{lewenstein:07, bloch:08,bloch:12,georgescu:14,gross:17,schaefer:20}. Recent highlights include the quantum simulation of correlations in Fermi-Hubbard systems~\cite{bergschneider:19} and the quantization of conductance through a quantum point contact~\cite{krinner:15,corman:19}. An important aspect underpinning all cold-atom-based quantum simulators is that their readout is single-shot images.
Experimental techniques for recording single-shot images include photoluminescence or indirect microscopy and tomography~\cite{Buecker:2009,Bakr:2009,Sherson:2010,Smith:2011,Estrecho:2018}. These readouts of ultracold atomic systems are projective measurements of the many-body wavefunction~\cite{javanainen:96,castin:97,dziarmaga:03,dagnino:09,sakmann:16,gajda:16,rakshit:17}. 
Ideally, such single-shot images -- random samples of the $N$-body density -- contain information about the position or momentum of every imaged atom and, thus, about  densities and correlation functions to all orders~\cite{schweigler:17,langen:15a,leveque:20}, quantum fluctuations~\cite{lin:20,chatterjee:18}, and full distribution functions~\cite{chatterjee:20}. 

A key goal is to have a quantum simulator whose readouts directly furnish information about complex correlation functions and distributions relevant, for instance, to our understanding of the nature of strongly correlated phases.  
Typically sought observables -- one- and two-body correlation functions -- are extracted by averaging a sufficiently large number of single-shot images. This averaging corresponds, formally, to the extraction of a marginal distribution via a trace operation. Such a trace operation might not preserve all the information about correlation functions. Reaching towards the goal to optimally readout quantum simulators, we propose to use artificial neural networks (ANNs) to optimally extract a wide range of pertinent observables from single-shot images.

The booming field of machine learning (ML) has led to many notable results concerning  the classification of phases of matter with simulated~\cite{wang:16, torlai:16, broecker:17, carrasquilla:17, carrasquilla_PRX:17, evert:17, wetzel:17, hu_PRE:17, zhang:17, schindler:17, costa:17, khatami:18, rao:18, wan:19} or experimental~\cite{bohrdt:19, rem:19, neugebauer:20} data.
ANNs were used to obtain ``neural-network quantum states'', a compressed yet accurate representation of many-body wavefunctions~\cite{carleo:17, gao:17, nomura:17, deng_PRX:17, deng:17, glasser:18, torlai:18, cai:18, chen:18, kaubruegger:18, carleo_exact:18, choo:18, vieijra:20}, and to devise efficient quantum state tomography and observable extraction in qubit systems~\cite{torlai:20, neugebauer:20,torlai:19, aolita:19}.
Applications of ANNs to single-shot images of ultracold atoms so far have focused on classifying phases of the doped Hubbard model~\cite{bohrdt:19}, detection of phase transitions of the Haldane and Bose-Hubbard models~\cite{rem:19}, and de-noising for single-exposure imaging~\cite{ness:20}. 
Hitherto, a demonstration that ANNs can be applied for the extraction of general observables including correlation functions from single-shot images of ultracold atoms is lacking.
% =================================

%%%%%%%%%%%%%%%%%%%%%%%%%%%%
\begin{figure}[h]
	\includegraphics[width=\columnwidth]{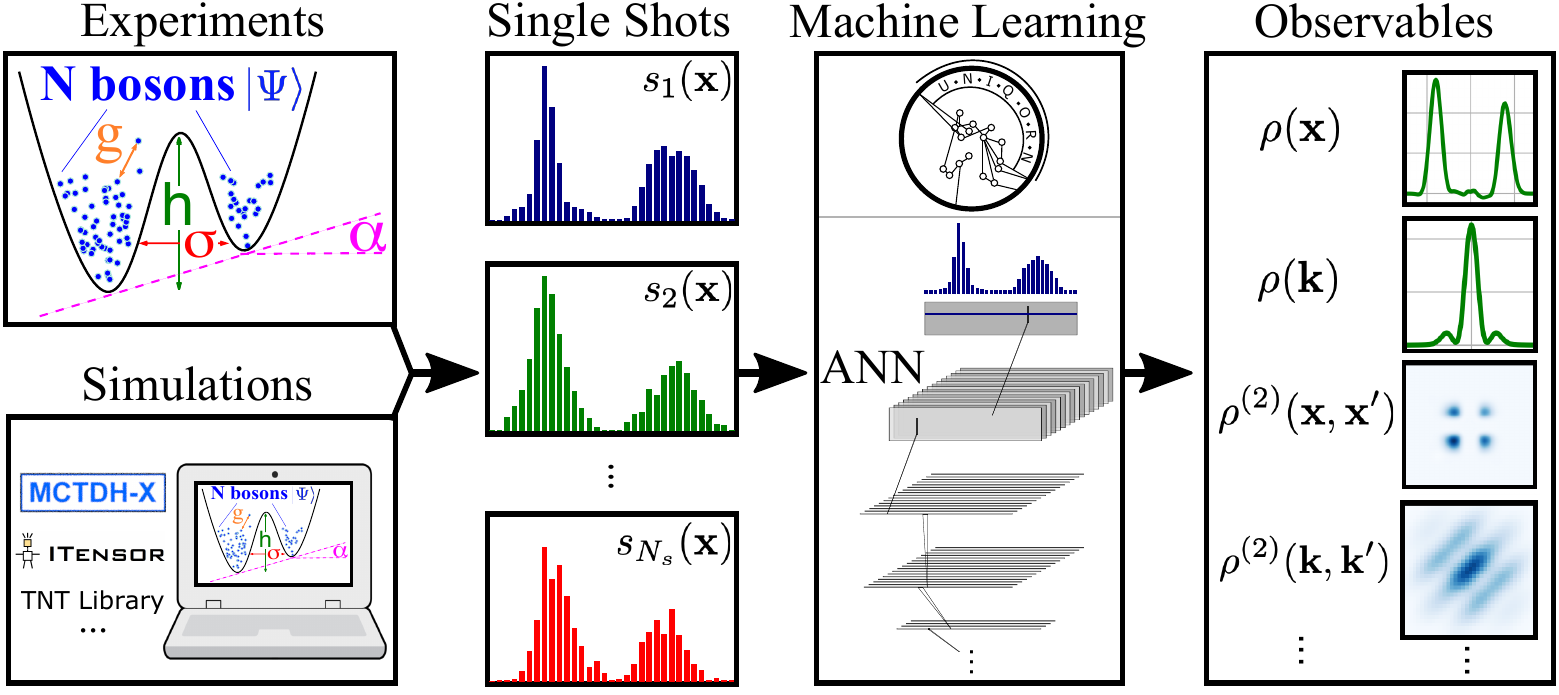}
	\caption{Machine learning observables from single-shot images of ultracold atomic systems. Single-shot images are measured experimentally or simulated numerically using MCTDH-X or other algorithms and used as input data in an ML task. Other options for wavefunction-based approaches that allow the simulation of single-shot measurements include ITensor~\cite{itensor}, the Tensor Network Theory (label TNT) Library~\cite{al-assam:17}, or the openMPS software~\cite{wall:12,jaschke:18}. The ANNs are trained on the simulated single-shot dataset obtained here for an $N$-boson tilted double-well system to output multiple observables.  
	The quantities ($g,h,\sigma,\alpha,N$) characterizing this double well are randomized. 
	}
	\label{fig:model}
\end{figure}
%%%%%%%%%%%%%%%%%%%%%%%%%%%%

%%%%%%%%%%%%%%%%%%%%%%%%%%%%%%%%%%%%%%
%%%% UNIQORN                      %%%%
%%%%%%%%%%%%%%%%%%%%%%%%%%%%%%%%%%%%%%
In this Letter, we devise, train, and apply ANNs to regress observables from simulated single-shot images of many-body systems of ultracold atoms [Fig.~\ref{fig:model}a)]. This ANN-based regression yields observables with far superior accuracy as compared to the standard averaging procedure.
As a key result of this Letter, we demonstrate an ANN-based reconstruction of observables from single-shot measurements that would traditionally require a change in the imaging setup: momentum-space observables can be found from real-space single-shot data and vice versa.
The ANN in use has a few convolutional and dense hidden layers, and is described in detail in Sec.~S1 of Supplementary Information (SI)~\footnote{Supplementary Information at [URL] discussing the used ML methods, MCTDH-X theory, ANN-based extraction of the particle number and the reduced one-body density matrix. The SI cites the additional Refs.~\cite{lecun:98,beck:00,zanghellini:03,alon:07_mix,bakr:09,buecker:09,sherson:10,smith:11,alon:12,miyagi:13,schmelcher:13,haxton.pra2:15,ioffe:15,goodfellow_book:16,klambauer:17,leveque:17,miyagi:17,schmelcher:17,leveque:18,lenail:19}}. 
The construction, training, and evaluation of ANN models is implemented using Tensorflow~\cite{abadi:15,chollet:15} and integrated in a flexible open-source toolkit: the Universal Neural-network Interface for Quantum Observable Readout from $N$-body wavefunctions, for short, UNIQOR$N$~\cite{UNIQORN_code}.

%%%%%%%%%%%%%%%%%%%%%%%%%%%%%%%%%%%%%%
%%%%   MODEL                      %%%%
%%%%%%%%%%%%%%%%%%%%%%%%%%%%%%%%%%%%%%

We aim to demonstrate the potential of our ML algorithm on a physical system that consists of $N$ bosons in a one-dimensional tilted double-well potential described by the Hamiltonian
\begin{align}
\mathcal{H} &= \int \mathrm{d} \mathbf{x} \: \hat{\Psi}^{\dagger}(\mathbf{x}) \left[ T(\mathbf{x}) + V(\mathbf{x}) \right] \hat{\Psi}(\mathbf{x}) \nonumber \\ 
& \quad + \frac{1}{2} \int \mathrm{d} \mathbf{x} \mathrm{d} \mathbf{x}' \:  \hat{\Psi}^{\dagger}(\mathbf{x}) \hat{\Psi}^{\dagger}(\mathbf{x}') W(\mathbf{x},\mathbf{x}') \hat{\Psi}(\mathbf{x}') \hat{\Psi}(\mathbf{x}), 
\label{eq:hamiltonian}
\end{align}
where $\hat{\Psi}(\mathbf{x}),\hat{\Psi}^{\dagger}(\mathbf{x})$ are bosonic field operators.
The one-body part of the Hamiltonian contains a kinetic energy term $T(\mathbf{x}) = -\frac{1}{2}  \partial_x^2$, a tilted one-body double-well potential $V(\mathbf{x}) =  \frac{1}{2} \mathbf{x}^2 + h \exp \left( -\frac{\mathbf{x}^2}{\sigma^2} \right) + \alpha \mathbf{x}$, modeled as a combination of an external harmonic confinement, a central Gaussian barrier of height $h$ and width $\sigma$, and a tilt of slope $\alpha$.
The particles interact via a two-body contact repulsion $W(\mathbf{x},\mathbf{x}') = g\delta(\mathbf{x} - \mathbf{x}')$, $g>0$ [Fig.~\ref{fig:model}b)].
All units are given in terms of the natural length scale $L=\sqrt{\hbar/m\omega}$ and energy scale $E=\hbar\omega$, where $m$ is the mass of the particles and $\omega$ is the external harmonic trapping frequency.

The tilted double-well system can be viewed as a minimal implementation of a quantum simulator of the two-site Hubbard model~\cite{murmann:15,bergschneider:19}, when the shape of the state within each well is ignored. The two lattice sites are then given by the two minima of the one-body potential, the hopping strength is controlled by the barrier height, and the on-site interaction and energy offset are derived from the interaction strength and the tilt, respectively. 
Despite their apparent simplicity, double wells feature a wealth of many-body properties: correlations~\cite{sakmann:08,sakmann:14,schweigler:17}, number fluctuations~\cite{sakmann:11b}, self-trapping and equilibration dynamics~\cite{sakmann:09} and are of contemporary experimental interest~\cite{buecker:11,langen:15a,schweigler:17,pigneur:18}. Therefore, double wells provide the perfect arena to implement and benchmark our ANN approach to extract observables.

%%%%%%%%%%%%%%%%%%%%%%%%%%%%
% X-->X FIG
% X-->X FIG
\begin{figure*}[!]
	\centering
	\includegraphics[width=\textwidth]{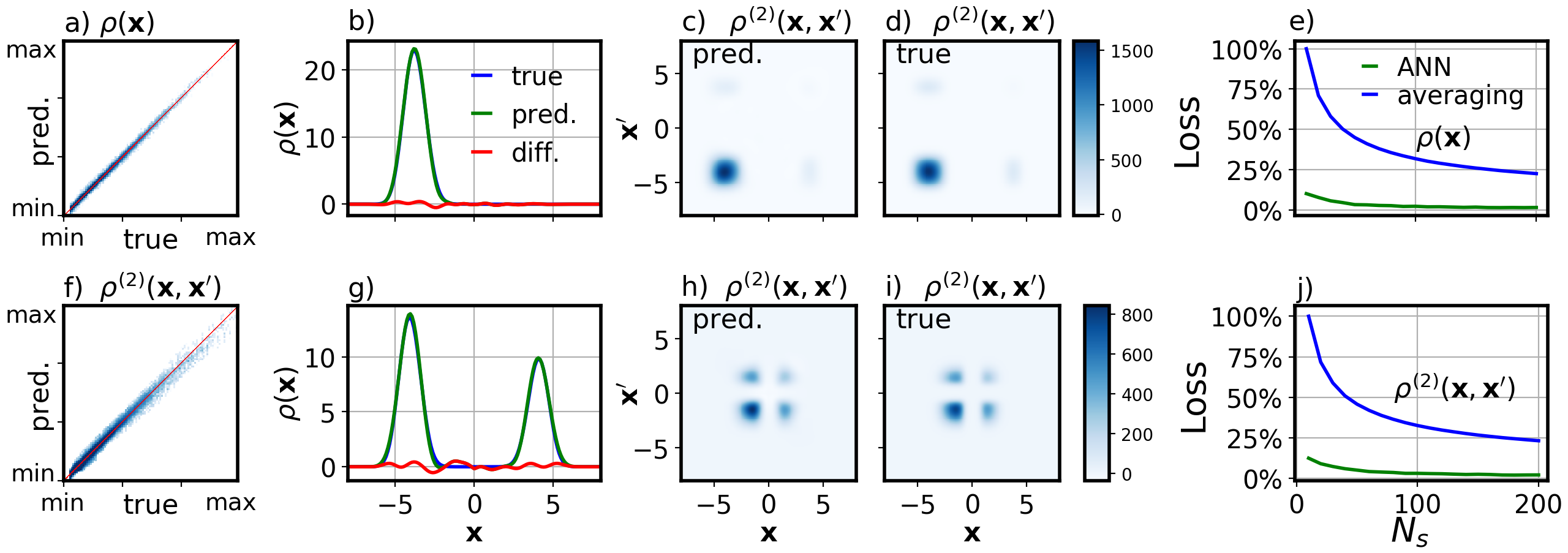}
	\caption{Regression of real-space 1BD, $\rho(\mathbf{x})$ [a),b),e),g)] and 2BD, $\rho^{(2)}(\mathbf{x},\mathbf{x}')$ [c),d),f),h),i),j)] from real-space single-shot images.
	a)-d),f)-i) Comparisons between predicted (label ``pred.'') and true (label ``true'') values from ANNs with $N_{s}=200$, for $\rho(\mathbf{x})$ (a),b),g)) and $\rho^{(2)}(\mathbf{x},\mathbf{x}')$ (c),d),f),h),i)).
Each dot in a),f) compares each regressed data point in all test set observables with its corresponding true value. The red line indicates perfect agreement for all ranges.
	In b),c),d), one specific condensed state is shown, while in g),h),i), one specific fragmented state is shown.
	e),j) The loss function measuring deviation of predictions from true values (Sec.~S1 in SI~\cite{Note1}) as a function of $N_{s}$ compared between the ANN approach and the averaging approach for e) $\rho(\mathbf{x})$ and j) $\rho^{(2)}(\mathbf{x},\mathbf{x}')$ ($100\%$ corresponds to the loss obtained for the averaging formula at $N_s=10$).
	}
	\label{fig:regression-real-space}
\end{figure*}
% X-->X FIG
% X-->X FIG
%%%%%%%%%%%%%%%%%%%%%%%%%%%%

%%%%%%%%%%%%%%%%%%%%%%%%%%%%%%%%%%%%%%
%%%% DATASET              		%%%%
%%%%%%%%%%%%%%%%%%%%%%%%%%%%%%%%%%%%%%
As the dataset~\cite{UNIQORN_dataset} for our ML tasks, we compute $3000$ ground-state wavefunctions of the double-well system described above. 
Each individual wavefunction  corresponds to a double-well system  whose  system parameters  barrier height $h\in \left[ 5, 25 \right]$, barrier width $\sigma \in \left[ 0, 3 \right]$, interactions $g\in \left[0.01, 0.2 \right]$, tilt $\alpha\in \left[0, 0.5 \right]$, and particle number $N\in \left[10, 100 \right]$ are generated randomly according to a uniform distribution within the corresponding intervals.
The ranges for the random parameters are tuned such that the obtained ground states span a range of physical phenomena that are expected for interacting ultracold bosons in the double-well potentials. 
Our dataset thus contains wavefunctions of both condensed and fragmented systems where the reduced one-body density matrix has a single~\cite{penrose:56} or several~\cite{spekkens:99,nozieres:82} macroscopic [$\mathcal{O}(N)$] eigenvalues, respectively. 

For each wavefunction $\Psi(\mathbf{x}_1,  \dots, \mathbf{x}_N)$ in our dataset of solutions of the Schr\"{o}dinger equation with the Hamiltonian~\eqref{eq:hamiltonian}, we generate $1000$ single-shot images as the input data for our ML tasks. These simulated single-shot images are random samples $(\tilde{\mathbf{x}}_1, \dots, \tilde{\mathbf{x}}_N)$ that are drawn from the $N$-particle density  $P(\mathbf{x}_1, \dots, \mathbf{x}_N) = \left|\Psi(\mathbf{x}_1, \dots, \mathbf{x}_N)\right|^2$, see Refs.~\cite{sakmann:16,lode:17}.
As the labels for our supervised ML regression tasks, we compute the one-body and two-body densities (1BDs and 2BDs, respectively) that we want to infer from the single-shot input data. This input data and its labels are computed in real and in momentum space and form our labeled dataset~\cite{UNIQORN_dataset}.
For convenience, we discuss the real-space data in the following; the discussion is identical for the momentum-space-case, replacing $\mathbf{x}\rightarrow \mathbf{k}$.
Since we have a range of different particle numbers, the shot-to-shot fluctuations in the single-shot images in our dataset stem from, both, the finite particle number and quantum fluctuations~\cite{tsatsos:17,chatterjee:20}. Hence, individual single-shot images in our dataset may be very different from their density labels. This deviation of single-shot images from the density is particularly pronounced for small particle numbers and/or strong quantum correlations~\cite{sakmann:16,lode:17}.

To solve the $N$-boson  Schr\"odinger equation for the ground state and to simulate the system's detection in single-shot images~\cite{lin:20,chatterjee:20,chatterjee:18,lode:17,tsatsos:17}, we use the multiconfigurational time-dependent Hartree method for indistinguishable particles (MCTDH-X)~\cite{alon.jcp:07,streltsov:07,alon:08,lode2:16,fasshauer:16,lin:20,ultracold,lode:20}. MCTDH-X employs a fully optimized basis set to obtain an accurate representation of the many-body state (Sec.~S2 in SI~\cite{Note1}). Generally, MCTDH-X can capture the physics of the system beyond the Hubbard model~\cite{mistakidis:14,mistakidis:15,neuhaus-steinmetz:17,mistakidis:17,dutta:19}.

%%% X--> X inferrence
%%% X--> X inferrence
%%% X--> X inferrence

Inspired by Ref.~\cite{bergschneider:19}, we study here the ANN-based extraction of 1BDs and 2BDs from single-shot images due to their simplicity and importance. 
Notably, the 2BD has an important role as the simplest indicator of correlation effects that underpins many-body phenomena beyond mean-field approaches~\cite{sakmann:08,sakmann:14}. 
We remark that our ANN-based approach works also for other observables like the particle number and the one-body reduced density matrix (Sec.~S3 and Sec.~S4 in SI~\cite{Note1}, respectively).

From the wavefunction $\vert \Psi \rangle$, the 1BD and 2BD can, respectively, be evaluated as 
\begin{eqnarray}
\rho(\mathbf{x}) &=& \langle \Psi \vert \hat{\Psi}^\dagger(\mathbf{x}) \hat{\Psi}(\mathbf{x}) \vert \Psi \rangle/N, \label{eq:rhox}\\ 
\rho^{(2)}(\mathbf{x},\mathbf{x}') &=& \langle \Psi \vert \hat{\Psi}^\dagger(\mathbf{x})  \hat{\Psi}^\dagger(\mathbf{x}') \hat{\Psi}(\mathbf{x}') \hat{\Psi}(\mathbf{x}) \vert \Psi \rangle. \label{eq:corr2}
\end{eqnarray}
Fig.~\ref{fig:regression-real-space} shows some results of our ANN-based extraction of the 1BD [Eq.~\eqref{eq:rhox}] and 2BD [Eq.~\eqref{eq:corr2}] in the top and bottom row, respectively, from a set of simulated real-space single-shot images $\lbrace s_i(\mathbf{x}); i=1,...,N_{s} \rbrace$ as a function of the number of images per input sample $N_{s}$.
The detailed description of the ANNs, including the regularizations for preventing overfitting, is shown in Sec.~S1 in SI~\cite{Note1}.

%%%%%%%%%%%%%%%%%%%%%%%%%%%%
% X-->K FIG
% X-->K FIG
\begin{figure*}[!]
	\centering
	\includegraphics[width=\textwidth]{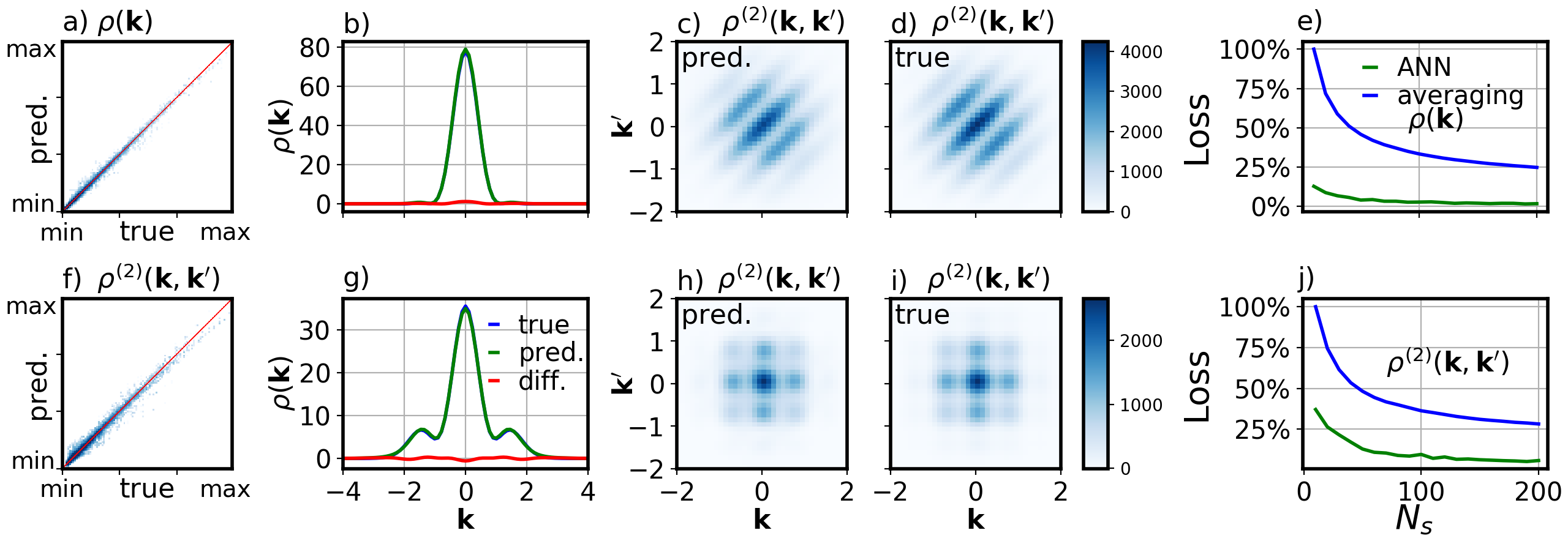}
	\caption{
		Regression of momentum-space 1BD $\rho(\mathbf{k})$ [a),b),e),g)] and 2BD $\rho^{(2)}(\mathbf{k},\mathbf{k}')$ [c),d),f),h),i),j)] from real-space single-shot images.
		a)-d),f)-i) Comparisons between predicted (label ``pred.'') and true (label ``true'') values from ANNs with $N_{s}=200$ for a),b),g) $\rho(\mathbf{k})$ and c),d),f),h),i) $\rho^{(2)}(\mathbf{k},\mathbf{k}')$. 
		In a),f), the results from the complete test set are shown, in b)-d), one specific condensed state is shown, while in g)-i), one specific fragmented state is shown.
	e),j) The loss function measuring deviation of predictions from true values (Sec.~S1 in SI~\cite{Note1}) as a function of $N_{s}$, compared between ANN and averaging approach for e) $\rho(\mathbf{k})$ and j) $\rho^{(2)}(\mathbf{k},\mathbf{k}')$ ($100\%$ corresponds to the loss obtained for the averaging formula at $N_s=10$). 
	See Sec.~S5 in SI~\cite{Note1} for complementary results on extracting real-space observables from momentum-space single shots $\lbrace s_i(\mathbf{k}), i=1,...,N_{s} \rbrace$.
	}
	\label{fig:regression-k-space}
\end{figure*}
% X-->K FIG
% X-->K FIG
%%%%%%%%%%%%%%%%%%%%%%%%%%%%

In comparison and in contrast to the ANN method, the 1BD and 2BD can also be approximately reconstructed by the averaging approach
\begin{eqnarray}
\rho(\mathbf{x}) &\approx& \frac{1}{N_{s}} \sum_{i=1}^{N_{s}} s_i(\mathbf{x}),
\label{eq:rho-formula}\\
\rho^{(2)}(\mathbf{x},\mathbf{x}') &\approx&  \frac{1}{N_{s}}  \sum_{i=1}^{N_{s}} s_i(\mathbf{x})\left[ s_i(\mathbf{x}') - \delta(\mathbf{x},\mathbf{x}') \right].
\label{eq:rho2-formula}
\end{eqnarray}
where $s_i(\mathbf{x})$ are the single-shot images.

In Fig.~\ref{fig:regression-real-space}, we illustrate the performance of the ANN-based approach with direct comparisons of the predicted and true values of $\rho$ and $\rho^{(2)}$.
Indeed, a very good agreement is seen between the predictions and the true values on the full test set [Fig.~\ref{fig:regression-real-space}~a),f)], spanning all physical regimes with very small (bottom left corner) to very large densities (upper right corner)~\footnote{Note that the larger values seem to rarefy simply because they typically correspond to states that are completely localized in one side of the double well, and are statistically less likely to appear for a randomly distributed range of parameters.}.
Furthermore, the predicted densities faithfully reproduce the corresponding true values for all different patterns of coherence.
This is exemplified for a condensed [Fig.~\ref{fig:regression-real-space}~b)-d)] and a fragmented [Fig.~\ref{fig:regression-real-space}~g)-h)] state. 
In the fragmented state, spatial correlations between the particles in distinct wells are present: $\rho^{(2)}(\mathbf{x},\mathbf{x}')$ is significant at $(\mathbf{x},\mathbf{x}')$ where $\mathbf{x}$ is in the vicinity of one of the minima of the double-well potential and $\mathbf{x}'$ in the vicinity of the other. For the condensed state, such correlations are absent.

To contextualize the performance of the observable reconstruction more concretely, we further explicitly compare the ANN-based accuracy by evaluating the loss function for its predictions on test data. Namely, we directly compare the discrepancy in the ANN-based approach with its counterpart for the averaging approach in Fig.~\ref{fig:regression-real-space}~e),j).
Strikingly, the ANN-based outperforms by more than one order of magnitude the averaging approach \emph{for all numbers of single-shot images per sample}, $N_{s}=10$.
This means that the ANN not only drastically reduces the number of single-shot images needed to achieve a certain accuracy -- the ANN even grants a better performance with that small number (\textit{e.g.} $N_{s}=10$) when compared to the averaging approach with a large number, i.e. $N_s=200$. Even larger numbers of samples are used in some experiments; for instance, $N_s \sim \mathcal{O}(10^3)$ in the experiments in Refs.~\cite{bergschneider:19,schweigler:17} on extracting correlations for ultracold atoms in double-well systems.
We infer that the ANN-based approach can harness the information about the state of the system $\vert \Psi \rangle$ embedded in the shot-to-shot fluctuations and needed for the construction of observables that bear correlation effects much more efficiently. We remark that the regression of momentum-space 1BD and 2BD from momentum-space single-shot images is performed equally well with the ANN-based approach, see Sec.~S5 and Fig.~S7 in SI~\cite{Note1}.

%%% X--> K inferrence
%%% X--> K inferrence
%%% X--> K inferrence

Motivated by these promising results for the cases where a benchmark with an averaging formula exists, we now apply ANNs to extract observables for which hitherto there exists no method of inference.
For this purpose, we consider the problem of inferring \emph{momentum-space} observables from \emph{real-space} single-shot images.
We thus feed sets of simulated real-space single-shot images, $\lbrace s_i(\mathbf{x})\rbrace$, to our ANNs as input data and train them to reconstruct the momentum-space densities $\rho(\mathbf{k})$ and $\rho^{(2)}(\mathbf{k},\mathbf{k}')$ [replacing $\mathbf{x} \rightarrow \mathbf{k}$ in Eq.~\eqref{eq:rhox} and Eq.~\eqref{eq:corr2}, respectively]. 

A conventional formula in the spirit of Eqs.~\eqref{eq:rho-formula} and \eqref{eq:rho2-formula} is unavailable for the inference of momentum-space observables from real-space single-shot images.
Nevertheless, we can reconstruct them from momentum-space single-shot images, $\lbrace s_i(\mathbf{k}); i=1,...,N_{s} \rbrace$, using Eqs.~\eqref{eq:rho-formula} and \eqref{eq:rho2-formula} after replacing $\mathbf{x}\rightarrow\mathbf{k}$. The loss function of this extraction approach can be used for comparison with the ANN loss function.

Fig.~\ref{fig:regression-k-space} shows the performance of the ANN-based extraction of momentum-space observables $\rho(\mathbf{k})$ and  $\rho^{(2)}(\mathbf{k},\mathbf{k}')$ from real-space single-shot images $\lbrace s_i(\mathbf{x})\rbrace$.
Again, the ANN-based extraction yields an outstanding accuracy. 
The loss function for $N_{s} \approx 200$ is drastically reduced with the ANN approach as compared to the averaging.
Additionally, the ANN-based loss function shows an equally good performance at low $N_{s} \approx 10$ as the conventional extraction approach at high $N_{s}\approx 200$.
From a visual inspection, we again find that all salient features of the 1BD and 2BD in momentum space are reproduced with large quantitative accuracy by the ANN [Fig.~\ref{fig:regression-k-space},b),g) and c),d),h),i), respectively]. 
Notably, the two-body densities are clearly similar to the fermionic results presented in Ref.~\cite{bergschneider:19}, confirming again the presence (absence) of correlations in a fragmented (condensed) state. 
The high accuracy of the ANN-based regression, even when compared to a direct averaging from momentum-space data, suggests that data obtained from in-situ imaging setup and processed with 
ANN architectures could be sufficient to extract reliable momentum-space observables.
This would render the potentially time-consuming reconfiguration of the imaging setup to a time-of-flight configuration unnecessary, and lead to an extraordinary reduction of lab resources.

%%%%%%%%%%%%%%%%%%%%%%%%%%%%%%%%%%%%%%%%%%%%%%
%%%% CONCLUSIONS                          %%%%
%%%%%%%%%%%%%%%%%%%%%%%%%%%%%%%%%%%%%%%%%%%%%%

We thus show that ANN-based observable extraction is a highly promising method for improving the performance of standard measurements in ultracold-atom-based experiments.  
The impact of our findings is far-reaching and multifaceted.
Information on a wide range of different states with various degrees of coherence can be retrieved reliably and rapidly.
The number of single-shot images and, therefore, the runtime of a costly experimental setup can be drastically reduced when using an ANN in comparison to when using the conventional averaging approach.
Moreover, we demonstrate how to extract momentum-space observables from in-situ single-shot images and vice versa without any tedious reconfiguration of the imaging setup and at a very high degree of accuracy.
Our results thus herald the potential of ANNs to obtain even more information from ultracold-atom-based quantum simulators. 

The next step is to obtain a first proof of concept by applying our ANN-based extraction to experimental single-shot images of single- and double-well setups~\cite{albiez:05,kierig:08,desbuquois:17,schweigler:17,pigneur:18,bergschneider:19,pruefer:20}. 
Further applications include the investigation of more involved cold-atom setups beyond single and double wells and tough-to-measure quantities like the potential~\cite{buettner:20} or higher-order densities and correlation functions~\cite{leveque:20}.

%%%%%%%%%%%%%%%%%%%%%%%%%%%%%%%%%%%%%%%%%%%%%%
%%%%               		       ACKNOWLEDGMENTS              				%%%%
%%%%%%%%%%%%%%%%%%%%%%%%%%%%%%%%%%%%%%%%%%%%%%

\begin{acknowledgements} 
This work has been supported by the Austrian Science Foundation (FWF) under grants P-32033-N32 and M-2653, the Swiss National Science Foundation (SNSF) and ETH grants, EPSRC Grants No. EP/P009565/1 and is partially funded by the European Research Council under the European Union's Seventh Framework Programme (FP7/2007-2013)/ERC Grant Agreement No. 319286 Q-MAC. 
Computation time on the Hazel Hen and Hawk clusters at the HLRS Stuttgart and on the ARCUS cluster of the University of Oxford, and support by the German Research Foundation (DFG) and the state of Baden-W\"urttemberg via the bwHPC grants no INST 40/467-1 FUGG (JUSTUS cluster), INST 39/963-1 FUGG (bwForCluster NEMO), and INST 37/935-1 FUGG (bwForCluster BinAC) is gratefully acknowledged.
\end{acknowledgements}
%%%%%%%%%%%%%%%%%%%%%%%%%%%%%%%%%%%%%%%%%%%%%%

%%%%%%%%%%%%%%%%%%%%%%%%%%%%%%%%%%%%%%%%%%%%%%
%%%%               				BIBLIOGRAPHY               					%%%%
%%%%%%%%%%%%%%%%%%%%%%%%%%%%%%%%%%%%%%%%%%%%%%
\bibliography{UNIQORN}
%%%%%%%%%%%%%%%%%%%%%%%%%%%%%%%%%%%%%%%%%%%%%%

\end{document}

\title{{\color{blue} Supplementary Material: \\
		 Optimized Observable Readout from Single-shot Images of Ultracold Atoms\\ via Machine Learning}}

\date{\today}

\author{Axel U. J. Lode}
\email{auj.lode@gmail.com}
\affiliation{Institute of Physics, Albert-Ludwig University of Freiburg, Hermann-Herder-Strasse 3, 79104 Freiburg, Germany}
\author{Rui Lin}
%\email{molignini@itp.phys.ethz.ch}
\affiliation{Institute for Theoretical Physics, ETH Z\"{u}rich, 8093 Zurich, Switzerland}
\author{Miriam B\"{u}ttner}
\affiliation{Institute of Physics, Albert-Ludwig University of Freiburg, Hermann-Herder-Strasse 3, 79104 Freiburg, Germany}
\author{Luca Papariello}
\affiliation{Research Studio Data Science, RSA FG, Thurngasse 8/16, 1090 Wien, Austria}
\author{Camille L\'{e}v\^{e}que}
\affiliation{Vienna Center for Quantum Science and Technology,
Atominstitut, TU Wien, Stadionallee 2, 1020 Vienna, Austria}
\affiliation{Wolfgang Pauli Institute c/o Faculty of Mathematics,
University of Vienna, Oskar-Morgenstern Platz 1, 1090 Vienna, Austria}
\author{R. Chitra}
\affiliation{Institute for Theoretical Physics, ETH Z\"{u}rich, 8093 Zurich, Switzerland}
\author{Marios C. Tsatsos}
\affiliation{Honest AI Ltd., 65 Goswell Road, London EC1V 7EN, United Kingdom}
\author{Dieter Jaksch}
\affiliation{Clarendon Laboratory, University of Oxford, Oxford OX1 3PU, United Kingdom}
\author{Paolo Molignini}
\email{phys2088@ox.ac.uk}
\affiliation{Clarendon Laboratory, University of Oxford, Oxford OX1 3PU, United Kingdom}

%%%%%%%%%%%%%%%%%%%%%%%%%%%%%%%%%%%%%%%%%%%%%%
%%%%               		   	          ABSTRACT              					%%%%
%%%%%%%%%%%%%%%%%%%%%%%%%%%%%%%%%%%%%%%%%%%%%%
\begin{abstract}
This supplementary information discusses our artificial neural networks and machine learning methods in Sec.~\ref{sec:ML}, the multiconfigurational time-dependent Hartree method for indistinguishable particles in Sec.~\ref{sec:MCTDHX}, and complementary machine learning results for extracting the particle number in Sec.~\ref{sec:NPAR}, the real- and momentum-space one-body density matrix in Sec.~\ref{sec:RHO1}, and the momentum-space and real-space densities from momentum-space single-shot images in Sec.~\ref{sec:KfromK} and Sec.~\ref{sec:XfromK}, respectively.
\end{abstract}
%%%%%%%%%%%%%%%%%%%%%%%%%%%%%%%%%%%%%%%%%%%%%%

\maketitle

%%%%%%%%%%%%%%%%%%%%%%%%%%%%%%%%%%%%%%%%%%%%%%
%%%%               			SUPPLEMENTARY              					%%%%
%%%%%%%%%%%%%%%%%%%%%%%%%%%%%%%%%%%%%%%%%%%%%%

%%%%%%%%%%%%%%%%%%%%%%%%%%%%%%%%%%%%%%
%%%%                   	                       METHODS                         		%%%%
%%%%%%%%%%%%%%%%%%%%%%%%%%%%%%%%%%%%%%

\section{Machine learning and neural network architectures} \label{sec:ML}

%%%%%%%%%%%%%%%%%%%%%%%%%%%%%%%%%%%%%%%%%%%%%%

We open with the details on the architectures for the artificial neural networks used to perform the regression of observables presented in the main text.
%%%%%%%%%%%%%%%%%%%%%%%%%%%%%%%%%%%%%%%%%%%%%%

For the machine learning (ML) task, we have trained a convolutional neural network (CNN) with three convolutional layers of different optimized lengths.
The exact architecture of the CNN is described in detail below.
Similar results were obtained also for a multi-layer perceptron (MLP), but we empirically found a superior performance for the CNN architecture.
This is consistent with better CNN performance in many other ML tasks in which the input data consists of images (see e.g. Refs.~\cite{lecun:98,carrasquilla:17,wan:19,levine:19})--the samples in our dataset can indeed be regarded as such.

The ANNs accept as input stacks of matrices of size $N_{x} \times N_{s}$, constructed by juxtaposing batches of $N_{s}$ single-shot images of length (number of points in $x$-direction) $N_{x}=256$.
The single-shot images were randomly re-shuffled to generate three times as many different images from the same dataset.

\subsection{Losses, data flow, and training}
To implement, train, validate, and test all ANNs we have employed the open-source TensorFlow library~\cite{chollet:15,abadi:15}.
As a loss function for all results besides those in Fig.~\ref{fig:regression-x-space}, we have selected the mean-squared error 
\begin{equation}
L_{\text{MSE}} = \frac{1}{N_{s}} \sum_{i=1}^{N_{s}} \left(y_i - \hat{y}_i \right)^2, \label{eq:MSE}
\end{equation}
where $y_i$ is the true value of the $i$-th data sample and $\hat{y}_i$ is its ANN-predicted counterpart.
For the regression of real-space observables from momentum-space single-shot data in Fig.~\ref{fig:regression-x-space}, we used a custom error function that is a sum of a $\log$-$\cosh$- and a mean absolute error measure:
\begin{eqnarray}
L_{\mathbf{k}\rightarrow \mathbf{x}} &=&  \frac{1}{N_{s}} \sum_{i=1}^{N_{s}} \log\left[\frac{1}{2}\left(\exp(y_i - \hat{y}_i) + \exp(- y_i + \hat{y}_i)\right)\right]\nonumber \\ 
&+& \frac{1}{N_{{s}}} \sum_{i=1}^{N_{s}} \left|y_i - \hat{y}_i \right|.
\label{eq:LossKX}
\end{eqnarray}
As the performance metrics we have used the mean-squared error loss in Eq.~\eqref{eq:MSE} (Eq.~\eqref{eq:LossKX} in the case of Fig.~\ref{fig:regression-x-space}) of the predictions of the ANN on a test set ($25\%$ our ground state wavefunctions) which we did not use for the training.
Where specified in the following subsections~\ref{subsubsec:mlp} and~\ref{subsubsec:cnn}, we augmented the loss function $L_\xi$ ($\xi=MSE$ or $\xi=\mathbf{k}\rightarrow \mathbf{x}$) with $L^2$-regularization,
\begin{equation}
L'_\xi = L_\xi + \sum_j r_j \sqrt{\sum_{i=1}^{N_j} \vert w_{ij} \vert^2 },
\end{equation}
and minimize the augmented loss $L'_\xi$ in the training. Here, $r_j$ is the regularization for the $j$-th layer to be regularized and $N_j$ is the number of weights $w_{ij}$ in that layer.
%%%%%%%%%%%%%%%%%%%%%%%%%%%%%%%%%%%%%%%%%%%%%%

The ANNs were trained via backpropagation with $75\%$ of the data samples in batches of size $N_B=10$ for a variable number of epochs $N_E\leq200$. 
The remaining $25\%$ of the data sets was used for the validation of the network on the test set.

We used the ``EarlyStopping'' and ``ReduceLRonPlateau'' callbacks to prevent overfitting. The EarlyStopping callback stops the training if the monitored performance metric -- in our case the losses on the validation set -- has stopped to improve.
The ReduceLRonPlateau callback is a scheduler for the so-called ``learning rate'', i.e., the step size of the minimization algorithm in use -- whenever the monitored performance metric (the validation loss) stops improving, the learning rate is decreased by a certain factor.
%

%%%%%%%%%%%%%%%%%%%%%%%%%%%%%%%%%%%%%%%%%%%%%%
\subsection{Multilayer perceptrons}
\label{subsubsec:mlp}

%%%%%%%%%%%%%%%%%%%%%%%%%%%%
\begin{figure}[!]
	\centering
	\includegraphics[width=\linewidth]{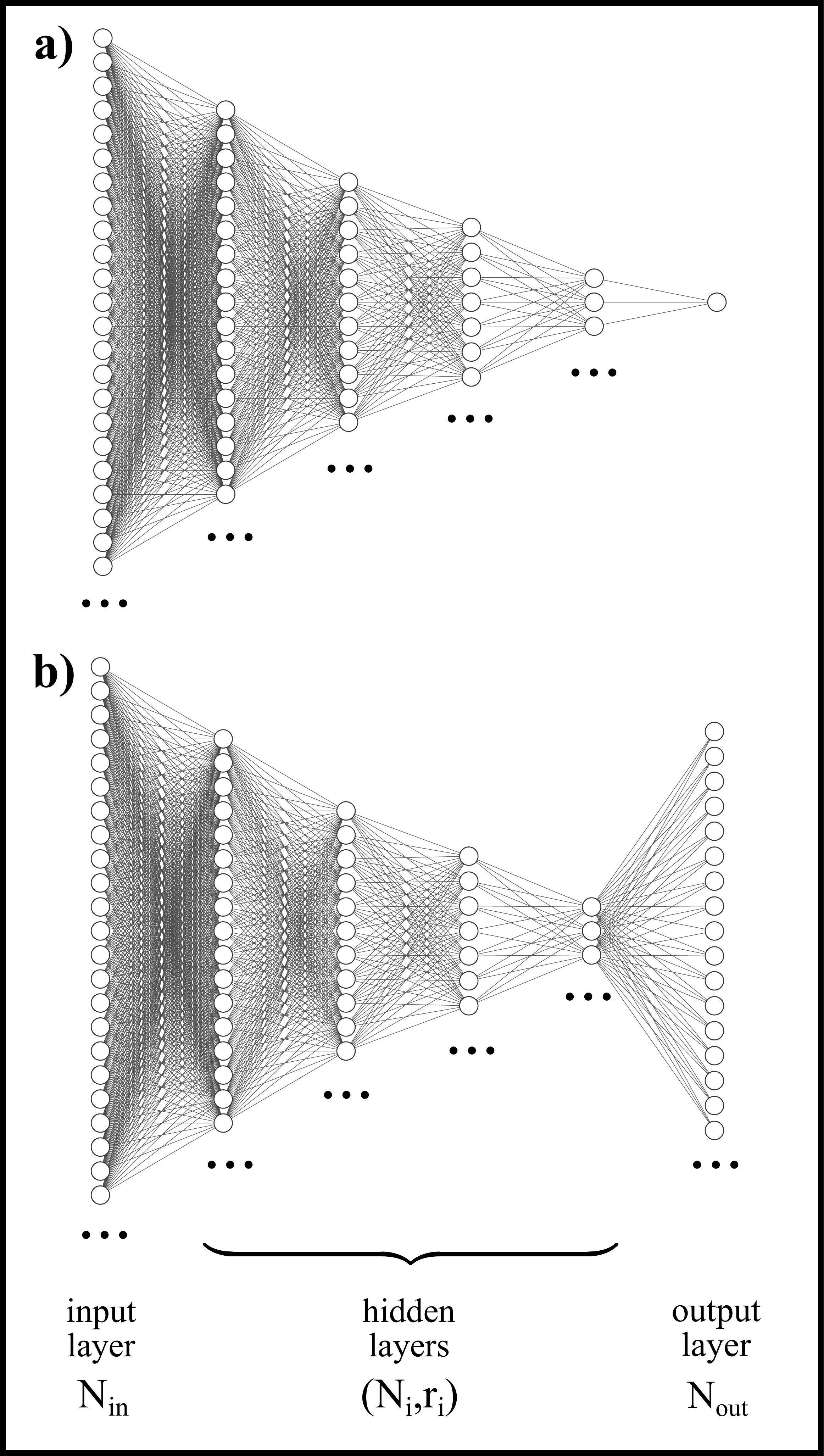}
	\caption{Illustration of the neural network architectures used for the MLPs: a) MLP for the regression of scalars such as particle number. b) MLP used for the regression of vectors such as density distributions. 
		For both cases an MLP with four hidden layers of decreasing lengths $N_1=512$, $N_2=256$, $N_3=128$, $N_4=64$ are used. 
		The parameters $r_i$ indicate the $L^2$ regularizations in each layer. For the results in Sec.~\ref{sec:NPAR}, we used $r_1=0.0,r_2=0.3,r_3=0.1,r_4=0.01$.
		Note that due to the large number of nodes in each layer, only a fraction of the nodes is depicted.
		This is also indicated by the three dots on bottom of each layer. Visualizations of the ANNs were done with the tool in Ref.~\cite{lenail:19}.}
	\label{fig:mlps}
\end{figure}
%%%%%%%%%%%%%%%%%%%%%%%%%%%%

%%%%%%%%%%%%%%%%%%%%%%%%%%%%%%%%%%%%%%%%%%%%%%
Multilayer perceptrons, also known as fully-connected feed-forward neural networks, belong to the most  paradigmatic deep learning models and consist of many artificial neurons connected into a network~\cite{goodfellow_book:16}. 
Neurons -- or nodes -- are collected in layers and MLPs are obtained by stacking layers together. 
The nodes in a layer are not connected among themselves, but each one is \emph{fully-connected} to those of neighboring layers (see Fig.~\ref{fig:mlps}).
Further, connections are directional and information flows from left to right (hence the name \emph{feed-forward}), i.e. from neurons in layer $i$ to neurons in layer $i+1$.

The MLP architectures used for our regression tasks are schematically depicted in Fig.~\ref{fig:mlps}. 
They consist of layers of fully-connected (dense) nodes, with four hidden layers of decreasing lengths.
For each layer we employed a scaled exponential linear unit (SeLU) as activation function, as they help with regularization~\cite{klambauer:17}.
Furthermore, we added kernel $L^2$-norm regularizers with the regularizations $r_i$ listed in Fig.~\ref{fig:mlps}.

%%%%%%%%%%%%%%%%%%%%%%%%%%%%%%%%%%%%%%%%%%%%%%
\subsection{Convolutional neural networks}
\label{subsubsec:cnn}
\begin{figure*}[!]
	\centering
	\includegraphics[width=\textwidth]{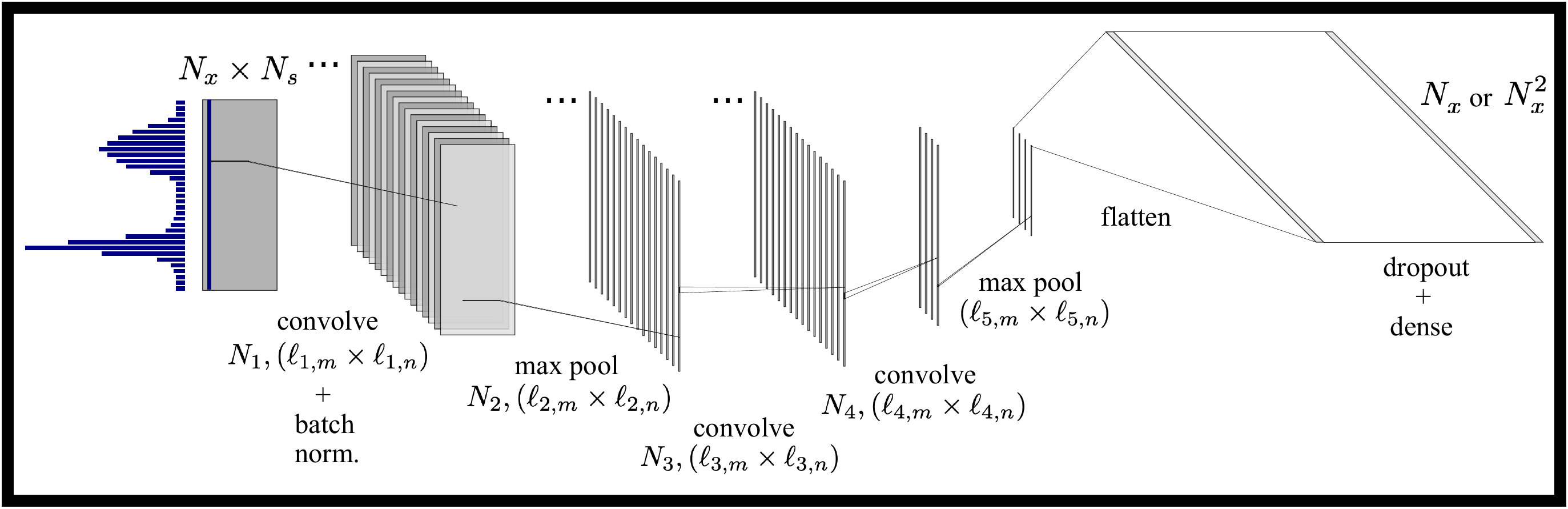}
	\caption{Illustration of the architecture used for the CNN for regression of observables.
		We begin by arranging together columns of different single-shot data (indicated by the blue column and the corresponding blue histogram) into a sample of length $N_{s}$. Our input data is thus ``images'' of size $256 \times N_s$. 
		The image is then processed with a series of different filters via convolution. The filters are either horizontally or vertically aligned (black lines in the grey rectangles), \textit{i.e.} they have shape $\ell_m \times 1$ or $1\times \ell_n$. The horizontally aligned filters thus are trained on the same coordinate across $\ell_m$ single shots, while the vertically aligned filters are trained on a single single-shot image on $\ell_n$ distinct coordinates.
		Note that due to the large number of filters in some layers, only a fraction of them is depicted (indicated by the dots next to the layer). To prevent overfitting, batch-normalization is performed after the first convolution~\cite{ioffe:15} and a dropout regularization is performed on the dense output layer (random neurons are dropped with a probability of $0.3$ to $0.5$). Visualizations of the ANNs were done with the tool in Ref.~\cite{lenail:19}.
}
	\label{fig:cnn}
\end{figure*}

%%%%%%%%%%%%%%%%%%%%%%%%%%%%%%%%%%%%%%%%%%%%%%
Convolutional neural networks have proven to be very successful in computer vision applications, which is where standard MLPs fall short. 
Their ability to retain information about locality and (approximate) translational invariance is the key ingredient that led CNNs to better deal with images~\cite{lecun:98, goodfellow_book:16}.
Another major advantage over MLPs lies in the fact that the number of trainable parameters does not scale directly with the size of the input data, which could otherwise lead to prohibitively large models.
Standard CNN architectures consist of an initial part of convolutional layers -- generally alternating with pooling layers (subsampling) -- followed by fully-connected layers (see Fig.~\ref{fig:cnn}) .

%%%%%%%%%%%%%%%%%%%%%%%%%%%%
%%%%%%%%%%%%%%%%%%%%%%%%%%%%

The CNN architecture used for our regression tasks is schematically pictured in Fig.~\ref{fig:cnn}. 
It consists of a sequence of horizontal and vertical filters that we train with the single-shot data arranged in ``images" of size $N_{x} \times N_{\text{s}}$.
In the first layer, we first pad the data and then convolve it with $N_1=256$ filters of size $\ell_{1,m} \times \ell_{1,n} = 1 \times 50$.
In the second layer, we act with a max pooling filter of size $\ell_{2,m} \times \ell_{2,n} = 1 \times 50$ to reduce the horizontal dimension.
In the third and fourth layers, we perform two other convolutions but with $N_3=256$ and $N_4=4$ ($N_4=16$ to obtain Fig.~\ref{fig:regression-x-space}) filters of size $\ell_{3,m} \times \ell_{3,n} = \ell_{4,m} \times \ell_{4,n} = 8 \times 1$ (``same'' padding).
We then max pool the data again with a filter of size $\ell_{5,m} \times \ell_{5,n} = 2 \times 2$ and flatten it out to obtain a single array of length $N_4 \left( \frac{N_x - (\ell_{3,m} + \ell_{4,m} - 2)}{\ell_{5,m}} \right) \frac{N_{\text{s}}}{\ell_{2,n} \ell_{5,n}}$, which is then mapped via a dense layer with dropout regularization (rates between $0.3$ and $0.5$) to the vectorized output layer containing the same number of neurons as there is values characterizing the observable that is to be regressed ($1$ for the particle number, $N_x$ for one-body densities, and $N_x^2$ for two-body densities and reduced one-body density matrices).
For all layers, we employed a SeLU activation function and a stride of $1$.

%%%%%%%%%%%%%%%%%%%%%%%%%%%%%%%%%%%%%%%%%%%%%%
\section{MCTDH-X and numerical details}\label{sec:MCTDHX}
%%%%%%%%%%%%%%%%%%%%%%%%%%%%%%%%%%%%%%%%%%%%%%
\subsection{MCTDH-X}

In this section we present the multiconfigurational time-dependent Hartree method for indistinguishable particles (MCTDH-X) software~\cite{lode2:16,fasshauer:16,lin:20,ultracold,lode:20}, which is used for the simulation of interacting ultracold bosons and the single-shot images in this work.

The MCTDH-X software is hosted at \url{http://ultracold.org} and is an implementation of the MCTDH-X theory~\cite{alon:08,alon.jcp:07,streltsov:07} and can solve the Schrödinger equation with many-body Hamiltonian, 
\begin{align}
	\mathcal{H} &= \int \mathrm{d} x \: \hat{\Psi}^{\dagger}(x) \left[ T(x) + V(x) \right] \hat{\Psi}(x) \nonumber \\ 
	& \quad + \frac{1}{2} \int \mathrm{d} x \mathrm{d} x' \:  \hat{\Psi}^{\dagger}(x) \hat{\Psi}^{\dagger}(x') W(x,x') \hat{\Psi}(x') \hat{\Psi}(x), \label{eq:HAM}
\end{align}
which consists of a kinetic term $T$, a one-body potential $V$, and a two-body interaction $W$.
For this purpose, MCTDH-X uses an adaptive ansatz for the bosonic wavefunction
\begin{eqnarray}
\begin{split}
\vert \Psi(t) \rangle &= \sum_{\vec{n}} C_{\vec{n}}(t) \vert \vec{n}; t \rangle; \;\; \vec{n}=\left(n_1,...,n_M\right)^T;\\
\vert \vec{n}; t \rangle &= \mathcal{N}  \prod_{i=1}^M \left[ \hat{b}_i^\dagger(t) \right]^{n_i} \vert \text{vac} \rangle; \\
\phi_j(\mathbf{x};t)&=\langle \mathbf{x} \vert \hat{b}_j (t) \vert 0 \rangle.  \label{eq:ansatz}
\end{split}
\end{eqnarray}
There are $M$ orthonormal time-dependent single-particle states, which are denoted as $\phi_j(\mathbf{x};t)$ and are annhihilated (created) by the operator $\hat{b}_j^{(\dagger)}$. These single-particle states form different symmetric configurations $\vert \vec{n}; t \rangle$, which are then superposed with time-dependent coefficients $C_{\vec{n}}(t)$ to form the full many-body wavefunction $\vert\Psi(t)\rangle$.
The normalization factor of the configurations is given by $\mathcal{N}=\left(\prod_{i=1}^{M} n_i!\right)^{-1/2}$. 

The evolution of the coefficients $C_{\vec{n}}(t)$ and the single-particle wavefunctions $\phi_j(\mathbf{x};t)$ follows a coupled set of first-order differential equations. These equations of motion and their solutions are further explained and discussed in detail in Refs.~\cite{alon:08,alon.jcp:07,lode:16,lode2:16,fasshauer:16,lode:20,streltsov:07}. 
Once the many-body wavefunction, i.e., the coefficients and single-particle wavefunctions, are obtained, they can be used to calculate various observables. These observables are later used as the data and labels in our training, validation, and test datasets.
We remark here, that MCTDH-X is a member of a large family of multiconfigurational methods, see Refs.~\cite{zanghellini:03,schmelcher:13,schmelcher:17,beck:00,haxton.pra2:15,alon:07_mix,alon:12,miyagi:13,miyagi:17,leveque:17,leveque:18}.

\subsection{Observables and single-shot images}

The most important quantities for our present work are single-shot images~\cite{sakmann:16,lode:17,chatterjee:18,chatterjee:20,tsatsos:17,gajda:16,rakshit:17}, which contain a lot of information about ultra-cold atom systems and are measured in experiments~\cite{bakr:09,buecker:09,sherson:10,smith:11}. The MCTDH-X software can be used to simulate the measurement of single-shot images from a many-body wavefunction. In real space, positions $(\mathbf{x}_1$, $\mathbf{x}_2  \dots \mathbf{x}_N)$ are drawn at random according to the probability distribution
\begin{eqnarray}
	P(\mathbf{x}_1,...,\mathbf{x}_N) = \left|\Psi(\mathbf{x}_1, \mathbf{x}_2, \dots, \mathbf{x}_N)\right|^2.
\end{eqnarray}
to obtain one single-shot image
\begin{eqnarray}
	s_i(\mathbf{x}) = \sum_k \delta(\mathbf{x}_k).
\end{eqnarray}

The same procedure can be performed in the momentum space by replacing $\mathbf{x}\rightarrow \mathbf{k}$. For every many-body wavefunction, we typically need roughly $1000$ such images to extract the various observables of interest, e.g. density distribution in Eqs.~(2) and (3) of the main text.

Another useful observable is the reduced one-body density matrix (1-RDM):
\begin{eqnarray}
	\rho^{(1)}(\mathbf{x},\mathbf{x}') &=& \frac{1}{N}\langle \Psi \vert \hat{\Psi}^\dagger(\mathbf{x}') \hat{\Psi}(\mathbf{x}) \vert \Psi \rangle \nonumber \\
	&=& \sum_{i=1}^M \rho_i \phi^{(\mathrm{NO}),*}_i(\mathbf{x}')\phi^{(\mathrm{NO})}_i(\mathbf{x}).\label{eq:RHO1X}
\end{eqnarray}
with $\sum_{i=1}^M\rho_i=1$ and $\rho_1\ge\rho_2\ge\dots\ge\rho_M$. 
In the second line of the above equation, we used the so-called natural representation and wrote the 1-RDM using its eigenvalues $\rho_i$ (natural occupations) and eigenfunctions $\phi_i^{(\mathrm{NO})}$ (natural orbitals).
A state is said to be condensed when only one natural orbital is macroscopically occupied $\rho_1\sim\mathcal{N}$~\cite{penrose:56}, and it is said to be fragmented when multiple eigenvalues are significant, $\rho_k\sim\mathcal{N}$ for more than a single $k$~\cite{nozieres:82,spekkens:99}. In the case of $M=2$ eigenvalues $\rho_i$, the fragmentation of a state can thus be represented by the first natural occupation $\rho_1$, because $N=\rho_1+\rho_2$ holds.

\subsection{Generation of numerical data}
To generate data and labels for the ANNs, we employ MCTDH-X to propagate a state in imaginary time to obtain the ground state of the many-body Hamiltonian, which is defined by Eq.~(1) of the main text and Eq.~\eqref{eq:HAM} above. The parameters of the Hamiltonian are chosen randomly according to a continuous and uniform distribution within the following intervals: particle numbers $N \in \left[10, 100 \right]$, barrier heights $h \in \left[ 5, 25 \right]$, barrier widths $\sigma \in \left[ 0, 3 \right]$, potential tilts $\alpha \in \left[0, 0.5 \right]$, and interactions $g  \in \left[0.01, 0.2 \right]$. 

In total, we randomly choose $3000$ different parameter sets and generate their corresponding ground state wavefunctions. In the simulations, $M=2$ orbitals are used, which is sufficient for capturing most features of the double-well potential and is economic computationally. These orbitals are represented by a primitive basis set of $256$ plane wave functions, corresponding to a grid with $N_x=256$ points that are equidistantly distributed in the interval $[-14,14]$, see Appendix B in Ref.~\cite{beck:00}.

For each ground state wavefunction, we have calculated $1000$ single-shot images~\cite{sakmann:16,lode:17} with a resolution of $N_x=256$ spatial points or momenta, which we then used as independent variables for our regression tasks. We have also calculated the true values (dependent variables) for the observables, such as fragmentation [$\rho_i$ in Eq.~\eqref{eq:RHO1X}], one-body and two-body densities, and reduced one-body density matrices which are used as labels for our ANNs.

%%%%%%%%%%%%%%%%%%%%%%%%%%%%%%%%%%%%%%%%%%%%%%
\section{Regression of particle number}\label{sec:NPAR}
%%%%%%%%%%%%%%%%%%%%%%%%%%%%%%%%%%%%%%%%%%%%%%

%%%%%%%%%%%%%%%%%%%%%%%%%%%%
\begin{figure}[!]
	\centering
	\includegraphics[width=\columnwidth]{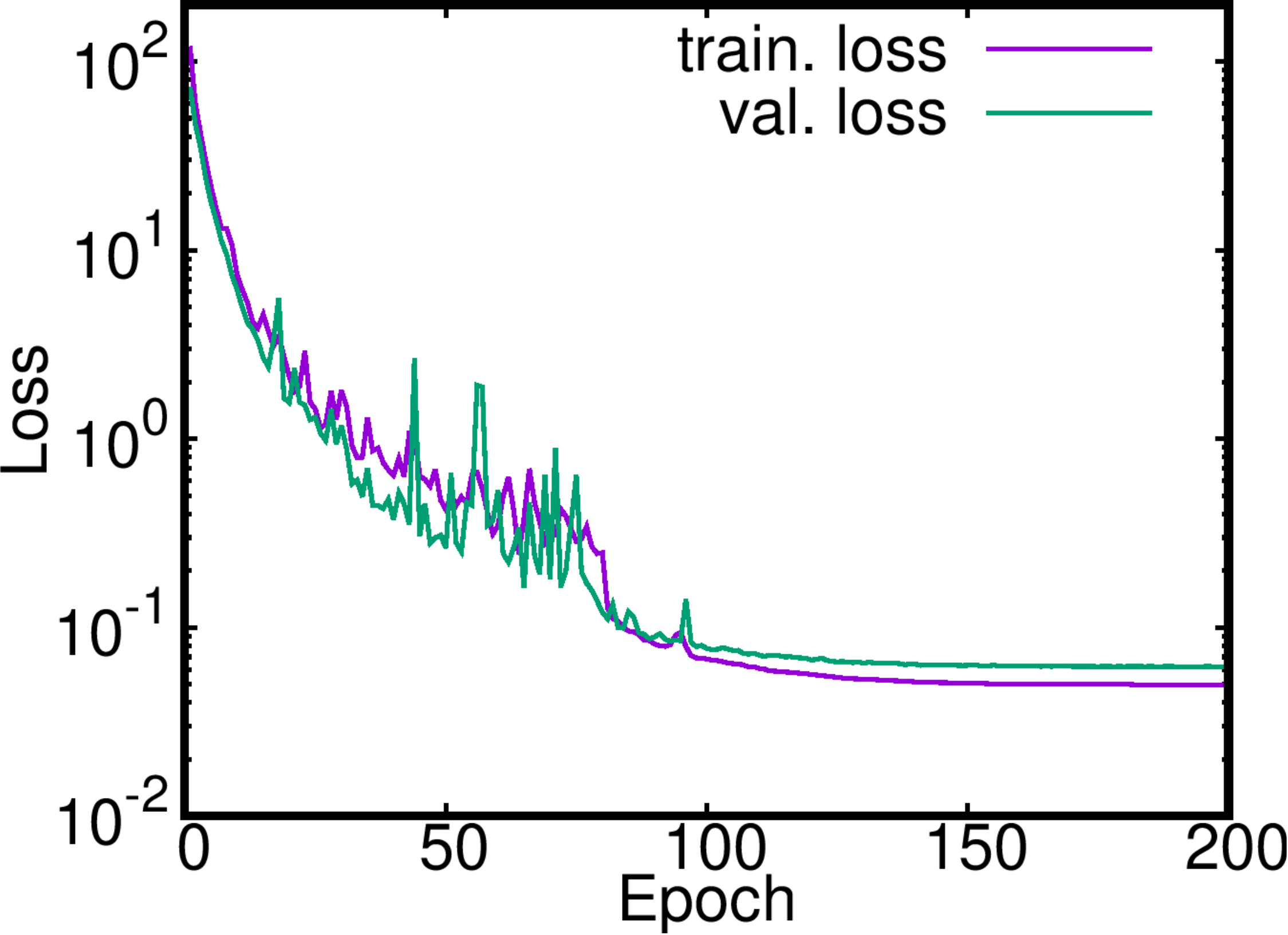}
	\caption{Training and validation mean-squared error of an MLP for the regression of the particle number for real-space single-shot data. Here, a dataset of $2000$ wavefunctions and $N_s=100$ single shots per input image were used.
	The  training (purple) and the validation (turquoise) loss  converge at $0.05$ and $0.07$ particles, respectively.
}
	\label{fig:REGR-NPAR_Loss}
\end{figure}
\begin{figure}[!]
	\centering
	\includegraphics[width=\columnwidth]{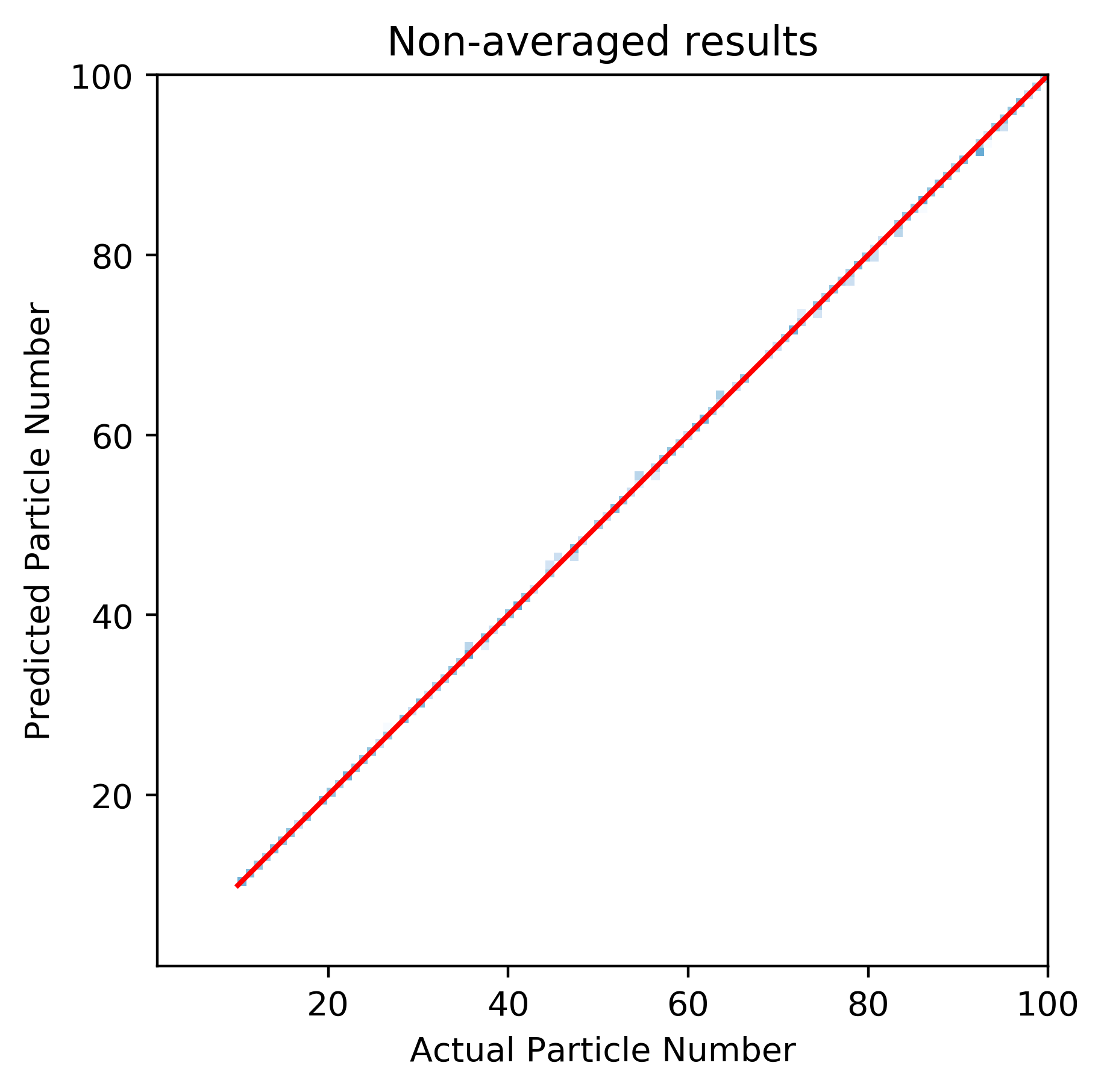}
	\caption{Performance of the MLP for predicting the particle number from the test set of real-space single-shot. The scatter plot is clustered very closely around the diagonal (red line) that marks the identity of predicted and true values.
	}
	\label{fig:REGR-NPAR_hist}
\end{figure}

%%%%%%%%%%%%%%%%%%%%%%%%%%%%
In this section we show the results obtained for the regression of particle number, for which we employed the MLP architecture trained over $N_E=200$ epochs.
The results are presented in Figs.~\ref{fig:REGR-NPAR_Loss} and \ref{fig:REGR-NPAR_hist}.
Fig.~\ref{fig:REGR-NPAR_Loss} shows the performance of the MLP in correctly regressing the data as a function of the epoch in the training and the validation process via the mean-squared error loss [cf. Eq.~\eqref{eq:MSE}]. 
From Fig.~\ref{fig:REGR-NPAR_Loss}, we can infer that the MLP quickly learns to correctly identify the number of particles in the training and validation sets ($1500$ wavefunctions).
Fig.~\ref{fig:REGR-NPAR_hist} shows a direct comparison of the predictions and the true values of the particle number on the test set.
Fig.~\ref{fig:REGR-NPAR_hist} demonstrates that the MLP's predictions are accurate throughout the whole range of particle numbers in the test set ($500$ wavefunctions).

 \section{Regression of real- and momentum-space one-body density matrix from real-space single-shot data}\label{sec:RHO1}
We now discuss the inference of the one-body reduced density matrix ($1$-RDM) from single-shot data. The $1$-RDM is defined in Eq.~\eqref{eq:RHO1X} in real space and as follows in momentum space: 
\begin{eqnarray}
\rho^{(1)}(\mathbf{k},\mathbf{k}') &=& \langle \Psi \vert \hat{\Psi}^\dagger(\mathbf{k})  \hat{\Psi}(\mathbf{k}') \vert \Psi \rangle. \label{eq:RHO1K}
\end{eqnarray}

\begin{figure*}[!]
	\centering
	\includegraphics[width=\textwidth]{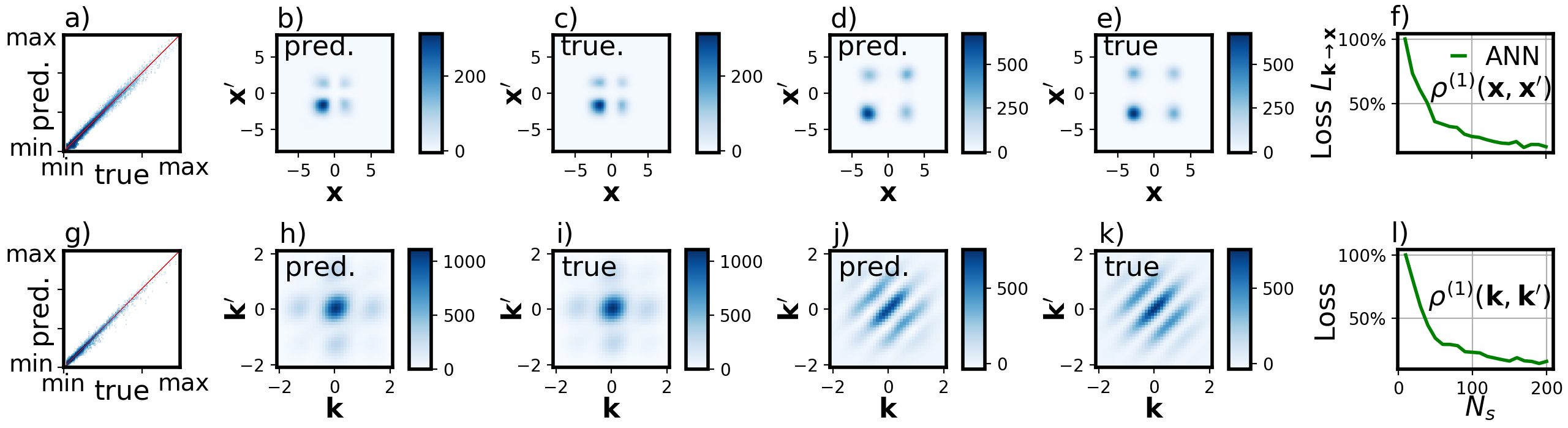}
	\caption{Regression of real-space and momentum-space reduced one-body density matrix, $\vert \rho^{(1)}(\mathbf{x},\mathbf{x}') \vert$ [a)--f)] and $\vert \rho^{(1)}(\mathbf{k},\mathbf{k}')\vert$ [g)--l)], respectively, from real-space single-shot images $\lbrace s_i(\mathbf{x}), i=1,...,N_{s} \rbrace$. 
		The ANN-predicted values (label ``pred.'') for $\vert \rho^{(1)}\vert$ are very close to the true (label ``true'') ones. For the case of $N_s=200$, panels a) and g) show a direct comparison of all the predicted values of the ANN for the test set and panels b)--e) and h)--k) show example comparisons of the predicted and the true values for the real-space and momentum-space one-body densities, respectively. $100\%$ in panels f),l) corresponds to the loss obtained at $N_s=10$.
	}
	\label{fig:RHO1}
\end{figure*}

In Fig.~\ref{fig:RHO1}, we plot the results of our ANN-based regression of $\vert \rho^{(1)}(\mathbf{x},\mathbf{x}') \vert$ and $\vert \rho^{(1)}(\mathbf{k},\mathbf{k}') \vert$.
The inference of the $1$-RDM works very accurately, both, in the real- and in the momentum-space case. This is a promising result, because the off-diagonal [$\mathbf{x}\neq \mathbf{x}'$ or $\mathbf{k}\neq \mathbf{k}'$] of the $1$-RDM is not directly measurable. We infer that the ANN learns a representation of the many-body state from the single-shot data that enables it to reconstruct the $1$-RDM. 

\begin{figure*}[!]
	\centering
	\includegraphics[width=\textwidth]{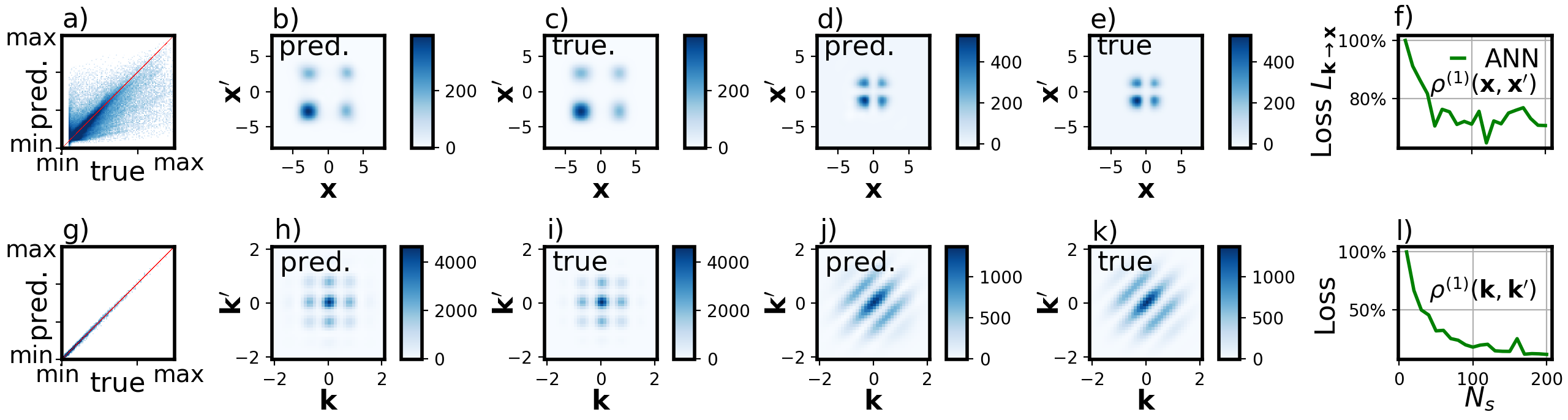}
	\caption{Regression of real-space and momentum-space reduced one-body density matrix, $\vert \rho^{(1)}(\mathbf{x},\mathbf{x}') \vert$ [a)--f)] and $\vert \rho^{(1)}(\mathbf{k},\mathbf{k}')\vert$ [g)--l)], respectively, from momentum-space single-shot images $\lbrace s_i(\mathbf{k}), i=1,...,N_{s} \rbrace$. 
		The ANN-predicted values (label ``pred.'') for $\vert \rho^{(1)}\vert$ are very close to the true (label ``true'') ones. 
		For the case of $N_s=200$, panels a) and g) show a direct comparison of all the predicted values of the ANN for the test set and panels b)--e) and h)--k) show example comparisons of the predicted and the true values for the real-space and momentum-space one-body densities, respectively. $100\%$ in panels f),l) corresponds to the loss obtained at $N_s=10$.
	}
	\label{fig:RHO1-2}
\end{figure*}

In Fig.~\ref{fig:RHO1-2}, we show the results for the regression of real- and momentum-space observables starting from momentum-space single-shot data.
The lower accuracy in the direct comparisons of true and predicted values for the regression of $\vert \rho^{(1)}(\mathbf{x},\mathbf{x}')|$ from momentum-space single shots [panel b)] is addressed in Sec.~\ref{sec:XfromK}.

%%%%%%%%%%%%%%%%%%%%%%%%%%%%%%%%%%%%%%%%%%%%%%
\section{Regression of momentum-space densities from momentum-space single-shot data}\label{sec:KfromK}
\begin{figure*}[!]
	\centering
	\includegraphics[width=\textwidth]{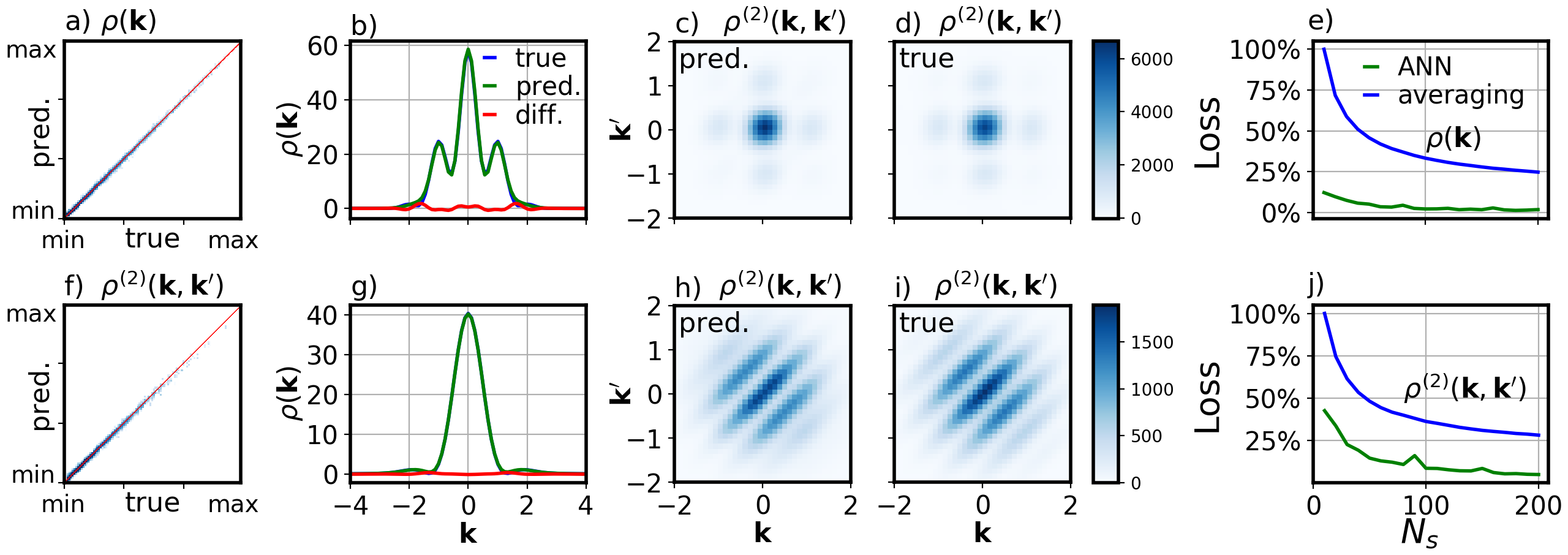}
	\caption{
		Regression of momentum-space one-body density, $\rho(\mathbf{k})$ [a),b),e),g)] and two-body density, $\rho^{(2)}(\mathbf{k},\mathbf{k}')$ [c),d),f),h),i),j)] from momentum-space single-shot images $\lbrace s_i(\mathbf{k}), i=1,...,N_{s} \rbrace$. 
		For the case of $N_s=200$, panels a) and f) show a direct comparison of all the predicted values of the ANN for the test set and panels b),g) and c),d),h),i) show example comparisons of the predicted and the true values for the one- and two-body density, respectively. 
		The ANN-predicted values (label ``pred.'') for $\rho(\mathbf{k})$ and $\rho^{(2)}(\mathbf{k},\mathbf{k}')$ are very close to the true ones (label ``true''). $100\%$ in panels e),j) corresponds to the loss obtained at $N_s=10$ with the averaging formula.
	}
	\label{fig:regression-k-space}
\end{figure*}

In this section, we provide results for the ANN-based extraction of momentum-space observables from momentum-space single-shot data, see Fig.~\ref{fig:regression-k-space}.
As for the regression of real-space observables from real-space single-shot data presented in the main text, the agreement between predicted and true one-body and two-body densities is excellent.
%%%%%%%%%%%%%%%%%%%%%%%%%%%%

%%%%%%%%%%%%%%%%%%%%%%%%%%%%%%%%%%%%%%%%%%%%%%
\section{Regression of real-space densities from momentum-space single-shot data}\label{sec:XfromK}
\begin{figure*}[htbp]
	\centering
	\includegraphics[width=\textwidth]{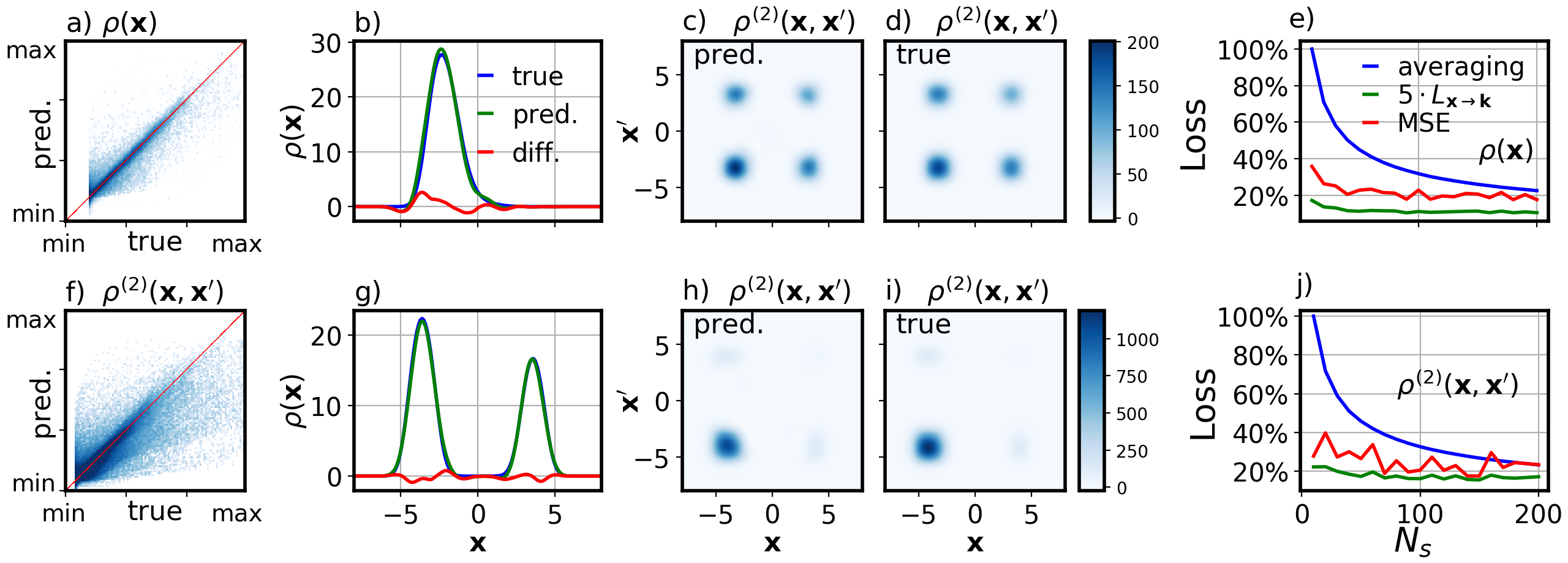}
	\caption{
		Regression of real-space one-body density, $\rho(\mathbf{x})$ [a),b),e),g)] and two-body density, $\rho^{(2)}(\mathbf{x},\mathbf{x}')$ [c),d),f),h),i),j)] from momentum-space single-shot images $\lbrace s_i(\mathbf{k}), i=1,...,N_{s} \rbrace$. 
		With ANNs, the achieved accuracy is much better for a small number $N_s$ of single-shot images per input sample for both $\rho(\mathbf{x})$ and $\rho^{(2)}(\mathbf{x},\mathbf{x}')$. 
		For convenience, we plot the mean-squared error [Eq.~\eqref{eq:MSE}] and the sum of the LogCosh and mean-absolute error [Eq.~\eqref{eq:LossKX}] in panels e) and j). $100\%$ in panels e),j) corresponds to the loss obtained at $N_s=10$ with the averaging formula.
		The optimization of the model was performed with minimizing the loss in Eq.~\eqref{eq:LossKX}.  
		At $N_s\approx 200$, the ANN-based inference of real-space observables from momentum-space single-shot data $\lbrace s_i(\mathbf{k}), i=1,...,N_{s} \rbrace$ becomes as accurate as the conventional averaging of real-space single-shot data $\lbrace s_i(\mathbf{x}), i=1,...,N_{s} \rbrace$ [e),j)]. The ANN-predicted values (label ``pred.'') for $\rho(\mathbf{k})$ and $\rho^{(2)}(\mathbf{k},\mathbf{k}')$ are very close to the true ones (label ``true''). For the case of $N_s=200$, panels a) and f) show a direct comparison of all the predicted values of the ANN for the test set and panels b),g) and c),d),h),i) show example comparisons of the predicted and the true values for the one- and two-body density, respectively. 
	}
	\label{fig:regression-x-space}
\end{figure*}

Here, we demonstrate that the ANN-based extraction of real-space observables from momentum-space single-shot data is practical, see Fig.~\ref{fig:regression-x-space}.

The overall agreement between the shape and absolute values of the predicted and the true one-body and two-body densities is very good.
In comparison to Fig.~3 in the main text, however, a wider spread of the prediction errors is observed [Fig.~\ref{fig:regression-x-space}a),f)].
We assume that the reason for this larger spread of the prediction errors is that a translation of a signal in real-space does not change the absolute value of the Fourier transform. The predicted densities are thus sometimes displaced with respect to the true ones, yielding a larger error.

%%%%%%%%%%%%%%%%%%%%%%%%%%%%

%%%%%%%%%%%%%%%%%%%%%%%%%%%%
% X-->K FIG
% X-->K FIG
% X-->K FIG
% X-->K FIG
%%%%%%%%%%%%%%%%%%%%%%%%%%%%

%%%%%%%%%%%%%%%%%%%%%%%%%%%%
% K-->K/X FIG
% K-->K/X FIG
% X-->K/X FIG
% X-->K/X FIG
%%%%%%%%%%%%%%%%%%%%%%%%%%%

% K-->K FIG
% K-->K FIG
% K-->K FIG
% K-->K FIG
%%%%%%%%%%%%%%%%%%%%%%%%%%%%		

% X-->K FIG
% X-->K FIG
\makeatletter\onecolumngrid@push\makeatother
% X-->K FIG
% X-->K FIG
%%%%%%%%%%%%%%%%%%%%%%%%%%%%
\clearpage
\makeatletter\onecolumngrid@pop\makeatother

%%%%%%%%%%%%%%%%%%%%%%%%%%%%%%%%%%%%%%%%%%%%%%
% force bib to the end
%%%%%%%%%%%%%%%%%%%%%%%%%%%%%%%%%%%%%%%%%%%%%%
%%%%               				BIBLIOGRAPHY               					%%%%
%%%%%%%%%%%%%%%%%%%%%%%%%%%%%%%%%%%%%%%%%%%%%%
\bibliography{UNIQORN}
%%%%%%%%%%%%%%%%%%%%%%%%%%%%%%%%%%%%%%%%%%%%%%